\documentclass[print, 
               notlongauthorlist 
              ]{nsr}

\volume{00}

\artnum{00}

\firstpage{1}

\datesubmitted{XX Month Year}
\datereceived{XX Month Year}
\daterevised{XX Month Year}
\dateaccepted{XX Month Year}
\datepublished{XX Month Year}

\doinum{doi/number}

\copyrightyear{2024}

\author[1]{M. G. Belyakova\footnote{Corresponding author e-mail address: \tt m.belyakova@lebedev.ru}$^{\orcidlink{0000-0000-0000-0000}}$}
\author[1]{R. B. Nevzorov\footnote{Corresponding author e-mail address: \tt nevzorovrb@lebedev.ru}$^{\orcidlink{0000-0001-7803-0507}}$}

\affil[1]{I.E.Tamm Theory Department, P.N. Lebedev Physical Institute of the Russian Academy of Sciences, Moscow, Russia}

\runauth{M. G. Belyakova and R. B. Nevzorov}

\title{Interactions of exotic neutralino dark matter with nucleons
in $U(1)$ extensions of the MSSM originating from $E_6$ GUTs}

\begin{document}


\maketitle

\begin{abstract}
\noindent
To ensure anomaly cancellation the $E_6$ inspired $U(1)$ extensions of the minimal
supersymmetric (SUSY) standard model (MSSM) involve extra exotic matter.
The lightest exotic neutralino in these models can be stable contributing to
the cold dark matter density. We consider the interactions of such neutralino
with nucleons within a specific extension of the MSSM with an additional $U(1)_N$ gauge
symmetry (SE$_6$SSM). The constraints on the couplings of this state, which
are set by the present experimental bounds caused by the direct detection experiments,
are examined. The obtained results can be generalised to other $E_6$ inspired
SUSY models with extra $U(1)$ gauge symmetry.
\end{abstract}




\section{Introduction}

Astrophysical and cosmological observations imply that non--luminous cold dark matter constitutes
about 20\%-25\% of the energy density of the Universe \cite{Ade:2015xua}. The nature of dark matter is unknown.
It cannot consist of any elementary particles discovered so far. Therefore the presence of dark matter is
widely considered as one of the strongest indications for new physics beyond the Standard Model (SM)
which describes well the experimental data measured in earth based experiments.

Models with softly broken supersymmetry (SUSY) remain the best motivated extensions of the SM.
The discovered SM--like Higgs scalar with mass around $125\,\mbox{GeV}$ is consistent with
the Minimal and Next-to-Minimal Supersymmetric Standard Models (MSSM and NMSSM). In these models
the electroweak (EW) scale is almost stabilized and the lightest neutralino is absolutely stable
if R--parity is preserved so that this state can play the role of dark matter. Because neutralinos
are heavy weakly interacting massive particles (WIMPs) they can explain the large scale structure of
the Universe \cite{Blumenthal:1984bp,Primack:2002th} providing the correct relic abundance of the cold dark matter \cite{Jungman:1995df}.
Within the MSSM the electroweak (EW) and strong gauge couplings extrapolated to high energies using the
renormalisation group equations (RGEs) converge to a common value near some high energy scale $M_X\sim 2\cdot 10^{16}\,\mbox{GeV}$
\cite{Ellis:1990wk,Langacker:1991an,Amaldi:1991cn,Anselmo:1991uu}.
This permits to embed the EW and strong interactions into Grand Unified Theories (GUTs) \cite{Georgi:1974sy},
which are based on $SU(5)$, $SO(10)$ or $E_6$ gauge groups, as well as into a ten-dimensional superstring theory
with $E_8\times E'_8$ gauge symmetry \cite{Green:1987sp}.

Around the GUT scale $M_X$ the gauge group $E_6$ (or $E_8$) reduces to $SO(10)\times U(1)_{\psi}$ (maximal subgroup of $E_6$)
with sequential breakdown of $SO(10)$ to $SU(5)\times U(1)_{\chi}$ and $SU(5)$ to the SM gauge group
$SU(3)_C\times SU(2)_W\times U(1)_Y$ (for reviews, see Refs.~\cite{E6-review,Langacker:2008yv,Nevzorov:2023dhd}). If the SUSY breaking
mechanism gives rise to the sparticle mass scale in the multi-TeV range then the breakdown of the
$U(1)_{\chi}\times U(1)_{\psi}$ symmetry to its discrete subgroup
\begin{equation}
U(1)_{\psi}\times U(1)_{\chi}\to P_M=(-1)^{3(B-L)}\,,
\label{1}
\end{equation}
where $B$ and $L$ are the baryon and lepton numbers, can result in a variety of SUSY
models including MSSM, NMSSM and their different modifications. The conservation of matter parity $P_M$ implies
that $R$--parity
\begin{equation}
Z_{2}^{R}=(-1)^{3(B-L)+2s}\,,
\label{2}
\end{equation}
where $s$ is the spin of the state, is also preserved.

The $U(1)$ extensions of the MSSM may arise in this
unification scheme when $U(1)_{\chi}\times U(1)_{\psi}$ symmetry is reduced to
\begin{equation}
U(1)'=U(1)_{\chi}\cos\theta_{E_6} + U(1)_{\psi}\sin\theta_{E_6}\,.
\label{3}
\end{equation}
If $\theta_{E_6}\ne 0$ and $\theta_{E_6}\ne\pi$ the anomalies in such $U(1)$ extensions of the MSSM get cancelled
if the particle spectrum involves complete representations of $E_6$. Consequently, one needs
to augment quarks and leptons of the SM by a number of exotics so that all these states
form three complete fundamental $27$ representations of $E_6$ at low energies.
These $27$--plets decompose under the $SU(5)\times U(1)_{\psi}\times U(1)_{\chi}$ subgroup of $E_6$ as follows:
$$
27_i \to
\left(10,\,\dfrac{1}{\sqrt{24}},\,-\dfrac{1}{\sqrt{40}}\right)_i
+\left(5^{*},\,\dfrac{1}{\sqrt{24}},\,\dfrac{3}{\sqrt{40}}\right)_i
+\left(5^{*},\,-\dfrac{2}{\sqrt{24}},\,-\dfrac{2}{\sqrt{40}}\right)_i
$$
\begin{equation}
+ \left(5,\,-\dfrac{2}{\sqrt{24}},\,\dfrac{2}{\sqrt{40}}\right)_i
+\left(1,\,\dfrac{4}{\sqrt{24}},\,0\right)_i
+\left(1,\,\dfrac{1}{\sqrt{24}},\,-\dfrac{5}{\sqrt{40}}\right)_i\,,
\label{4}
\end{equation}
The quantities in brackets are the $SU(5)$ representation as well as extra $U(1)_{\psi}$ and $U(1)_{\chi}$ charges,
while $i=1,2,3$ is a family index. An ordinary SM family, which includes the doublets of left--handed
leptons $L_i$ and quarks $Q_i$, right--handed charged leptons, up-- and down--quarks ($e^c_i, u^c_i$ and $d^c_i$),
is assigned to $\left(10,\,\dfrac{1}{\sqrt{24}},\,-\dfrac{1}{\sqrt{40}}\right)_i$ + $\left(5^{*},\,\dfrac{1}{\sqrt{24}},\,\dfrac{3}{\sqrt{40}}\right)_i$.
The last term in Eq.~(\ref{4}) corresponds to the right-handed neutrinos $N^c_i$.
The next-to-last term in Eq.~(\ref{4}) can be identified with extra SM-singlet fields $S_i$ with non-zero $U(1)_{\psi}$
charges. The pairs of $SU(2)_W$--doublets ($H^d_{i}$ and $H^u_{i}$) which are contained in
$\left(5^{*},\,-\dfrac{2}{\sqrt{24}},\,-\dfrac{2}{\sqrt{40}}\right)_i$ and $\left(5,\,-\dfrac{2}{\sqrt{24}},\,\dfrac{2}{\sqrt{40}}\right)_i$
have the quantum numbers of Higgs doublets (Higgs--like doublets) so that they form either Higgs or exotic Higgs $SU(2)_W$ multiplets.
The colour triplet components of these $SU(5)$ multiplets are associated with the exotic quarks $\overline{D}_i$ and $D_i$
with electric charges $+ 1/3$ and $-1/3$. The $B-L$ charges of these exotic quarks are twice larger than that of ordinary ones, i.e. $\left(\pm\dfrac{2}{3}\right)$. Thus they can be either leptoquarks or diquarks.

The $E_6$ inspired $U(1)$ extensions of the MSSM were extensively studied over the years. Different aspects of phenomenology of
the exotic quarks/squarks and $Z'$ boson were considered in Refs.~\cite{Kang:2007ib,Ali:2023kss,Accomando:2010fz}.
The implications of these models were explored for neutrino physics \cite{Kang:2004ix,Ma:1995xk}
and models explaining the hierarchy of fermion masses  \cite{Stech:2008wd}, EW baryogenesis \cite{Ma:2000jf,Kang:2004pp},
leptogenesis \cite{Nevzorov:2017gir,Nevzorov:2023scg,Hambye:2000bn,King:2008qb}, muon anomalous magnetic moment \cite{Grifols:1986vr,Morris:1987fm},
CP-violation in the Higgs sector \cite{Ham:2008fx} and EW symmetry
breaking \cite{Langacker:1998tc,Cvetic:1996mf,Cvetic:1995rj,Cvetic:1997ky,Suematsu:1994qm,Daikoku:2000ep,Keith:1997zb},
lepton flavour violating processes like $\mu\to e\gamma$ \cite{Suematsu:1997qt},
electric dipole moment of electron \cite{Suematsu:1997tv} and tau lepton \cite{GutierrezRodriguez:2006hb}.
The neutralino sector in such models was studied in \cite{Keith:1997zb,Suematsu:1997qt,Suematsu:1997tv,GutierrezRodriguez:2006hb,Suematsu:1997au,Keith:1996fv,Hesselbach:2001ri,Barger:2005hb,Choi:2006fz,
Barger:2007nv,Gherghetta:1996yr,E6neutralino-higgs}.
In the $E_6$ inspired $U(1)$-extended SUSY models the upper bound on the SM--like Higgs mass and
the Higgs sector were analysed in Refs.~\cite{Daikoku:2000ep}, \cite{E6neutralino-higgs,E6-higgs,King:2005jy,King:2005my,King:2006vu}.

Here we explore the interactions of dark matter particles with nucleons in the framework of specific $E_6$ inspired SUSY model
with an additional $U(1)_{N}$ gauge symmetry associated with $\theta_{E_6}=\arctan\sqrt{15}$ in Eq.~(\ref{3})
\cite{King:2005jy,King:2005my} (for recent review, see Ref.~\cite{King:2020ldn}). This exceptional supersymmetric standard model
(E$_6$SSM) implies that near the GUT scale $U(1)_{\psi}$ and $U(1)_{\chi}$ are reduced to
\begin{equation}
U(1)_{\psi}\times U(1)_{\chi}\to U(1)_{N}\times P_M\,.
\label{5}
\end{equation}
Only in this case the right--handed neutrinos have zero charges so that
they can be extremely heavy \cite{King:2005jy, King:2005my}. Although we restrict our analysis here to a particular value of
$\theta_{E_6}$ our results can be easily generalised to other $E_6$ inspired $U(1)$ extensions of the MSSM
in which matter parity is preserved.

In all $E_6$ inspired $U(1)$-extended SUSY models extra exotic matter may lead to rapid proton decay and non-diagonal
flavor transitions. A set of discrete symmetries permits to suppress the corresponding operators \cite{King:2005jy,King:2005my}.
In this paper we consider a variant of the E$_6$SSM (SE$_6$SSM) \cite{Nevzorov:2012hs,Athron:2014pua,Athron:2015vxg,Athron:2016gor}
in which the most dangerous baryon and lepton number violating operators as well as tree-level flavor-changing transitions
are forbidden by a single discrete $\tilde{Z}^{H}_2$ symmetry. Two discrete symmetries $\tilde{Z}^{H}_2$ and $P_M$
give rise to at least two stable states, i.e. gravitino and the lightest exotic neutralino, that may compose cold
dark matter density. In the previous article \cite{Nevzorov:2022zns} it was pointed out that the dark matter-nucleon
scattering cross section can be sufficiently strongly suppressed. Here we compare the computed values of this cross section
with the corresponding experimental bounds and identify the restrictions on the parameter space of the SE$_6$SSM
caused by the latest results of the LUX--ZEPLIN (LZ) experiment \cite{LZ:2024zvo}.

The layout of this article is as follows. In section 2 we briefly review the SUSY model under consideration.
In section 3 the constraints on the SE$_6$SSM parameter space, which come from the direct detection experiments, are considered.
Section 4 concludes the paper.

\section{Exotic neutralino dark matter}

Different modifications of the E$_6$SSM were considered in Refs.~
\cite{King:2005jy,King:2005my,Nevzorov:2012hs,Athron:2014pua,Howl:2007hq,Howl:2007zi,Howl:2008xz,Howl:2009ds,
Athron:2010zz,Hall:2011zq,Callaghan:2012rv,Callaghan:2013kaa,King:2016wep,Nevzorov:2018leq,Khalil:2020syr}.
To ensure the cancellation of anomalies below the GUT scale $M_X$ the matter content of the simplest variant
of the E$_6$SSM involves three $27$--plets as well as a pair of $SU(2)_W$ doublets, i.e. $L_4$ and $\overline{L}_4$ \cite{King:2020ldn}.
Because $L_4$ and $\overline{L}_4$ come from extra $27^{\prime}$ and $\overline{27}^{\prime}$, they carry opposite
$SU(3)_C \times SU(2)_W \times U(1)_Y \times U(1)_N$ quantum numbers and gauge anomalies still cancel.
In the simplest variant of the E$_6$SSM the supermultiplets $L_4$ and $\overline{L}_4$ facilitate the gauge coupling
unification near the scale $M_X$ \cite{King:2007uj}. In the vicinity of the quasi-fixed point the upper bound on the mass of the lightest
Higgs boson within this variant of the E$_6$SSM was analysed in~\cite{Nevzorov:2013ixa}. Such quasi-fixed point appears as an intersection
of the quasi-fixed and invariant lines~\cite{Nevzorov:2001vj,Nevzorov:2002ub,Froggatt:2007qp}.
Extra exotic states in the E$_6$SSM can give rise to distinctive LHC signatures
\cite{King:2005jy,King:2005my,Howl:2007zi,Athron:2010zz,King:2006vu,King:2006rh,Belyaev:2012si,Belyaev:2012jz}
and non-standard Higgs decays \cite{Athron:2014pua,Nevzorov:2013tta,Hall:2010ix,Hall:2010ny,Nevzorov:2014sha,Nevzorov:2020jdq,Athron:2016usd,Nevzorov:2015iya}.

Using approach proposed in~\cite{Hesselbach:2007ta,Hesselbach:2007te}, it was found that in the E$_6$SSM
the lightest supersymmetric particle (LSP), i.e. the lightest $R$--parity odd state,
has to be lighter than $60-65~\mbox{GeV}$~\cite{Hall:2010ix}. Such LSPs may account for some of the cold dark matter
density if their masses are close to half the $Z$ boson mass $M_Z/2$ \cite{Hall:2010ix}. Nevertheless in this
part of the E$_6$SSM parameter space the SM-like Higgs scalar decays predominantly into a pair of LSPs
while other branching ratios of the Higgs boson are rather small. LHC experiments ruled out such scenario.
The simplest phenomenologically viable scenario implies that LSPs in the E$_6$SSM are considerably lighter than $1\,\mbox{eV}$
forming hot dark matter in our Universe. They give only a minor contribution to the dark matter density.
The presence of very light neutral fermions may result in interesting implications for the neutrino physics \cite{Frere:1996gb}.

\begin{table}[!hbt]
\centering
\caption{The $U(1)_Y$ and $U(1)_{N}$ charges of the SE$_6$SSM supermultiplets.}\label{tab:tab1}
\begin{tabular}{ccccccccccccccc}
\toprule
                             & $Q_i$ & $u^c_i$ & $d^c_i$ & $L_i, L_4$ & $e^c_i$ & $N^c_i$ & $S_i, S$ & $H^u_{\alpha}, H_u$ &
                               $H^d_{\alpha}, H_d$ & $D_i$ & $\overline{D}_i$ & $\overline{L}_4$ & $\overline{S}$ & $\phi_i, \phi$\\
\midrule
 $\sqrt{\frac{5}{3}}Q^{Y}_i$ & $\frac{1}{6}$ & $-\frac{2}{3}$ & $\frac{1}{3}$ & $-\frac{1}{2}$ & $1$ & $0$ & $0$ & $\frac{1}{2}$
                             & $-\frac{1}{2}$ & $-\frac{1}{3}$ & $\frac{1}{3}$ & $\frac{1}{2}$ & $0$ & $0$\\
$\sqrt{{40}}Q^{N}_i$         & $1$ & $1$ & $2$ & $2$ & $1$ & $0$ & $5$ & $-2$ & $-3$ & $-2$ & $-3$ & $-2$ & $-5$ & $0$ \\
\bottomrule
\end{tabular}
\end{table}

In addition to three $27$--plets, $L_4$ and $\overline{L}_4$ the field content of the SE$_6$SSM also includes
a pair of superfields $S$ and $\overline{S}$ arising from another $\tilde{27}+\overline{\tilde{27}}$ representations
as well as four $E_6$ singlet superfields $\phi$ and $\phi_i$ ($i=1,2,3$). Extra superfields permit to avoid the
appearance of extremely light neutral fermions in the particle spectrum. The SE$_6$SSM matter content can originate from
the $E_6$ orbifold GUT model in six dimensions \cite{Nevzorov:2012hs}. The $U(1)_Y$ and $U(1)_{N}$ charges of all matter
supermultiplets in the SE$_6$SSM are given in Table~\ref{tab:tab1}. The supermultiplets $\phi$, $L_4$, $\overline{L}_4$,
$S$, $\overline{S}$ and one pair of the Higgs--like doublets (say $H_d\equiv H^d_{3}$ and $H_u\equiv H^u_{3}$)
are required to be even under the discrete $\tilde{Z}^{H}_2$ symmetry while the remaining supermultiplets have to be
$\tilde{Z}^{H}_2$ odd. Neglecting all suppressed non-renormalisable interactions, the SE$_6$SSM superpotential
can be written as
\begin{equation}
W_{\rm SE_6SSM} = \lambda S (H_u H_d) - \sigma \phi S \overline{S} +
\dfrac{\kappa}{3}\phi^3+\dfrac{\mu}{2}\phi^2+\Lambda\phi
+ \mu_L L_4\overline{L}_4+ \tilde{\sigma} \phi L_4\overline{L}_4 + W_{IH}
\label{6}
\end{equation}
$$
+ \kappa_{ij} S (D_{i} \overline{D}_{j}) + g^D_{ij} (Q_i L_4) \overline{D}_j+ h^E_{i\alpha} e^c_{i} (H^d_{\alpha} L_4)
+ g_{ij} \phi_i \overline{L}_4 L_j + W_N + W_{\rm MSSM}\,,
$$
\begin{equation}
W_{IH} = \tilde{M}_{ij} \phi_i \phi_j + \tilde{\kappa}_{ij} \phi \phi_i \phi_j
+ \tilde{\lambda}_{ij} \overline{S} \phi_i S_j  + \lambda_{\alpha\beta} S (H^d_{\alpha} H^u_{\beta})
+ \tilde{f}_{i\alpha} S_{i} (H^d_{\alpha} H_u) + f_{i\alpha} S_{i} (H_d H^u_{\alpha})\,,
\label{7}
\end{equation}
\begin{equation}
W_N =  \frac{1}{2} M_{ij} N_i^c N_j^c + \frac{1}{2} \eta_{ij} \phi N_i^c N_j^c + \tilde{h}_{ij} N_i^c (H_u L_j) +
h_{i\alpha}  N_i^c (H^u_{\alpha} L_4)\,,
\label{8}
\end{equation}
\begin{equation}
W_{\rm MSSM} = y^U_{ij} (Q_i H_u) u^c_j + y^D_{ij} (Q_i H_d) d^c_j + y^L_{ij} (L_i H_d) e^c_j\,.
\label{9}
\end{equation}
In Eq.~(\ref{6})--(\ref{9}) $\alpha, \beta =1,2$ and $i, j =1,2,3$.
When $M_{ij}=\eta_{ij}=0$ the expressions (\ref{6})--(\ref{9}) are valid for any $E_6$ inspired $U(1)$ extension of the MSSM
(with $\theta_{E_6}\ne 0$ and $\theta_{E_6}\ne\pi$) if $\tilde{Z}^{H}_2$ and $P_M$ are preserved. The $\tilde{Z}^{H}_2$ symmetry ensures that
the SE$_6$SSM superpotential does not contain any operators that lead to rapid proton decay. The $SU(2)_W$--doublets
$H_d$ and $H_u$ play the role of the MSSM Higgs supermultiplets. Because of the $\tilde{Z}^{H}_2$ symmetry the
down-type quarks and charged leptons couple to just $H_d$ whereas the up-type quarks couple to $H_u$ only. As a consequence
the non-diagonal flavour transitions are suppressed at tree-level. The Yukawa couplings of $L_4$ to $Q_i$ and $\overline{D}_j$,
which are allowed by the $\tilde{Z}^{H}_2$ symmetry, permit the lightest exotic quarks to decay within a reasonable time.

The $U(1)_N$ gauge symmetry is spontaneously broken in the SE$_6$SSM by the vacuum expectation values (VEVs) of
the superfields $S$ and $\overline{S}$. In the limit $\sigma\to 0$ the corresponding part of the scalar potential
is given by
\begin{equation}
V_S(S,\,\overline{S}) = m^2_S |S|^2 + m^2_{\overline{S}} |\overline{S}|^2
+\dfrac{Q_S^2 g^{\prime \, 2}_1}{2}\left(|S|^2-|\overline{S}|^2\right)^2\,,
\label{10}
\end{equation}
where $m_S^2$ and $m^2_{\overline{S}}$ are the soft SUSY breaking masses while
$g^{\prime}_1$ and $Q_S$ are the $U(1)_N$ gauge coupling and the $U(1)_N$ charge of $S$.
When $(m^2_S + m^2_{\overline{S}})<0$ the minimum of the potential (\ref{10}) is attained
for $\langle S \rangle = \langle \overline{S} \rangle \to\infty$.
This run-away direction is stabilized for non-zero values of coupling $\sigma$.
If $\sigma\ll 1$ then the VEVs
\begin{equation}
\langle \phi \rangle \sim \langle S \rangle \simeq \langle \overline{S} \rangle \sim \dfrac{M_S}{\sigma}\,,
\label{11}
\end{equation}
tend to be much larger than the sparticle mass scale $M_S$. The exotic states and $Z'$ boson are rather heavy
in this case.

For practical purposes, it is convenient to introduce $Z_{2}^{E}$ symmetry which is defined as
$\tilde{Z}^{H}_2 = P_{M}\times Z_{2}^{E}$ \cite{Nevzorov:2012hs}. The transformation properties of
the SE$_6$SSM supermultiplets under the $P_{M}$, $Z_{2}^{E}$ and $\tilde{Z}^{H}_2$ symmetries are
presented in Table~\ref{tab:tab2}. Since $\tilde{Z}^{H}_2$ and $P_{M}$ are exact the $Z_{2}^{E}$ symmetry
and $R$--parity are conserved. Therefore the lightest $R$--parity odd state should be stable.
Here we assume that gravitino is the lightest SUSY particle in the spectrum and can give a substantial
contribution to the cold dark matter density. Because gravitino is even under the discrete $Z^{E}_2$
symmetry the lightest $Z_{2}^{E}$ odd state has to be stable as well \cite{Nevzorov:2012hs}.

\begin{table}[!hbt]
\centering
\caption{Transformation properties of the SE$_6$SSM supermultiplets under the discrete symmetries.}\label{tab:tab2}
\begin{tabular}{ccccc}
\toprule
                              & ~~~~$Q_i, u^c_i, d^c_i, L_i, e^c_i, N^c_i$~~~~
                              & ~~~~$\overline{D}_i, D_i, H^d_{\alpha}, H^u_{\alpha}, S_{i}, \phi_i$~~~~
                              & ~~~~$H_d, H_u, S, \overline{S}, \phi$~~~~ & ~~~~$L_4, \overline{L}_4$~~~~\\
\midrule
$\tilde{Z}^{H}_2$             &  $-$              & $-$                   & $+$       & $+$ \\
$P_{M}$                       &  $-$              & $+$                   & $+$       & $-$ \\
$Z_{2}^{E}$                   &  $+$              & $-$                   & $+$       & $-$ \\
\bottomrule
\end{tabular}
\end{table}

The phenomenologically acceptable scenarios with gravitino LSP imply that all unstable particles
decay before Big Bang Nucleosynthesis (BBN). For the sparticle mass scale $M_S$ in the multi-TeV range
this requirement can be satisfied if gravitino mass $m _{3/2}$ is lower than $10\,\mbox{GeV}$ \cite{Nevzorov:2022zns}.
The contribution of such gravitinos to the cold dark matter density is determined by the reheating
temperature $T_R$ \cite{Bolz:2000fu,Eberl:2020fml}
\begin{equation}
\Omega_{3/2} h^2 \sim 0.27 \left(\dfrac{T_R}{10^8 GeV}\right) \left(\dfrac{1\,\mbox{GeV}}{m _{3/2}}\right) \left(\dfrac{M_{\tilde{g}}}{1\,\mbox{TeV}}\right)^2\,.
\label{12}
\end{equation}
where $M_{\tilde{g}}$ is a gluino mass. Because $\Omega_{3/2} h^2 \le 0.12$ \cite{Ade:2015xua},
for $M_{\tilde{g}}\gtrsim 3\,\mbox{TeV}$ and $m _{3/2}\simeq 1\,\mbox{GeV}$ one finds an upper limit
$T_R\lesssim 10^{6-7}\,\mbox{GeV}$ \cite{Hook:2018sai}. Nevertheless even for so low $T_R$
the appropriate baryon asymmetry can be generated within the SE$_6$SSM via the out--of equilibrium
decays of the lightest right--handed neutrino/sneutrino \cite{Nevzorov:2017gir}.

The part of the SE$_6$SSM superpotential $W_{IH}$ describes the interactions of the
$\tilde{Z}_2^H$ even supermultiplets $\phi$, $S$, $\overline{S}$, $H_u$ and $H_d$
with $\phi_i$, $S_i$, $H^u_{\alpha}$ and $H^d_{\alpha}$ which do not develop VEVs.
The fermion components of these $\tilde{Z}_2^H$ odd supermultiplets compose the exotic
neutralino and chargino states. The signatures associated with such exotic
neutralino states were studied in Refs.~\cite{Khalil:2021afa,Khalil:2021tpz}.
If all components of $\phi_i$ are much heavier than the fermions and bosons from
$H^u_{\alpha}$, $H^d_{\alpha}$ and $S_i$ the superfields $\phi_i$ can be integrated out
and $W_{IH}$ reduces to
\begin{equation}
\begin{array}{c}
W_{IH} \to \widetilde{W}_{IH}\simeq -\widetilde{\mu}_{ij} S_i S_j + \lambda_{\alpha\beta} S (H^d_{\alpha} H^u_{\beta})
+ \tilde{f}_{i\alpha} S_{i} (H^d_{\alpha} H_u) + f_{i\alpha} S_{i} (H_d H^u_{\alpha})+...\,.
\end{array}
\label{13}
\end{equation}
Here and further we use the basis in which $\lambda_{\alpha\beta}=\lambda_{\alpha\alpha}\,\delta_{\alpha\beta}$
and $\widetilde{\mu}_{ij}=\widetilde{\mu}_{i}\,\delta_{ij}$.

As mentioned before, the lightest exotic state with $Z_{2}^{E}=-1$ has to be stable. In this paper it is assumed that
such state is mostly formed by the neutral fermion components $\tilde{H}^{u0}_1$ and $\tilde{H}^{d0}_1$ of the
supermultiplets $H^u_{1}$ and $H^d_{1}$. We explore the scenarios in which
all sparticles except gravitino as well as all other exotic states are rather heavy, i.e. they are considerably
heavier than $1\,\mbox{TeV}$. It is also expected that the supermultiplets $H^u_{1}$ and $H^d_{1}$ mainly interact with
$S_1$, $H_u$ and $H_d$. To simplify the analysis all other couplings of these $SU(2)_W$--doublets are set to be
negligibly small. In this approximation the masses of the lightest exotic neutralinos are eigenvalues of the
mass matrix \cite{Nevzorov:2022zns}:
\begin{equation}
M^{ab}=-
\left(
\begin{array}{ccc}
0                                           & \mu_{11}                            & \dfrac{\tilde{f}_{11}}{\sqrt{2}} v_2 \\[3mm]
\mu_{11}                                    & 0                                   & \dfrac{f_{11}}{\sqrt{2}} v_1 \\[0mm]
\dfrac{\tilde{f}_{11}}{\sqrt{2}} v_2        & \dfrac{f_{11}}{\sqrt{2}} v_1        & \widetilde{\mu}_1             \\
\end{array}
\right)\,,
\label{14}
\end{equation}
where~ $v_1$~ and~ $v_2$~ are~ the~ VEVs~ of~ $H_d$~ and~ $H_u$,~ i.e.~ $\langle H_u \rangle = v_2/\sqrt{2}$,~ $\langle H_d \rangle = v_1/\sqrt{2}$~
and\\ $v=\sqrt{v_1^2+v_2^2} \approx 246\,\mbox{GeV}$. In Eq.~(\ref{14}) $\mu_{11}=\lambda_{11} \langle S \rangle$.

The perturbation theory method allows to diagonalise the mass matrix (\ref{14}) if $|\widetilde{\mu}_1|$ is much larger than
$v$ and $|\mu_{11}|$ (see, for example, \cite{Kovalenko:1998dc,Nevzorov:2000uv,Nevzorov:2001um}).
For $\mu_1 > 0$ and $\mu_{11} > 0$ one obtains
\begin{equation}
m_{\chi_1} \simeq \mu_{11} - \Delta_1\,,\qquad\quad m_{\chi_2} \simeq \mu_{11} + \Delta_2\,,\qquad\quad
m_{\chi_3} \simeq \widetilde{\mu}_1 + \Delta_1 + \Delta_2\,,
\label{15}
\end{equation}
$$
\Delta_1 \simeq \dfrac{(\tilde{f}_{11} v\sin\beta + f_{11} v\cos\beta)^2}{4(\widetilde{\mu}_1 - \mu_{11})}\,,\qquad\qquad
\Delta_2 \simeq \dfrac{(\tilde{f}_{11} v\sin\beta - f_{11} v\cos\beta)^2}{4(\widetilde{\mu}_1 + \mu_{11})}\,,
$$
where $\tan\beta=v_2/v_1$. Eqs.~(\ref{15}) indicate that the masses of the lightest and next-to-lightest exotic neutralinos
($m_{\chi_2}$ and $m_{\chi_1}$) are determined by $\mu_{11}$. In our analysis we require that $m_{\chi_2}-m_{\chi_1}> 200\,\mbox{MeV}$
so that the next-to-lightest exotic neutralino $\chi_2$ decays before BBN.

Using the approximate formula
\begin{equation}
\Omega_{\tilde{H}} h^2 \simeq 0.1 \, \left(\dfrac{\mu_{11}}{1\,\mbox{TeV}}\right)^2
\label{16}
\end{equation}
one can estimate the contribution of the lightest exotic neutralino $\chi_1$ to the dark matter density.
Eq.~(\ref{16}) was derived within the MSSM in the case when dark matter is formed by the Higgsino states
\cite{Arkani-Hamed:2006wnf,Chalons:2012xf}. Planck observations indicate that $(\Omega h^2)_{\text{exp}} = 0.1188 \pm 0.0010$ \cite{Ade:2015xua}.
Thus in the phenomenologically viable scenarios $\mu_{11}$ has to be lower than $1.1\,\mbox{TeV}$.
For $|\mu_{11}| < 1.1\,\mbox{TeV}$ gravitino may account for some part of the dark matter density.

\section{Direct detection constraints}

Now let us consider the interactions of dark matter states with the baryons within the SE$_6$SSM.
It is well known that gravitino with masses of a few GeV has negligibly small couplings to the SM particles.
Therefore the interactions of dark matter particles with nucleons are defined by the couplings of the stable
lightest exotic neutralino $\chi_1$. The two lightest exotic neutralinos are linear combinations of
$\tilde{H}^{d0}_1$, $\tilde{H}^{u0}_1$ and the fermion component $\tilde{S}_1$ of the superfield $S_1$, i.e.
\begin{equation}
\chi_{\alpha} = N_{\alpha}^1 \tilde{H}^{d0}_1 + N_{\alpha}^2 \tilde{H}^{u0}_1 + N_{\alpha}^3 \tilde{S}_1\,,
\label{17}
\end{equation}
where $\alpha=1,2$. The exotic neutralino mixing matrix $N^a_i$ ($a,i=1,2,3$) is defined by:
\begin{equation}
N_i^a M^{ab} N_j^b = m_i \delta_{ij}, \qquad\mbox{ no sum on } i.
\label{18}
\end{equation}
The two lightest exotic neutralinos mostly interact with the $Z$-boson and Higgs states.
The masses of all Higgs bosons except the SM-like one are set to be equal to $10\,\mbox{TeV}$.
As a consequence their contribution to the scattering of the lightest exotic neutralino on nuclei
can be ignored. The part of the Lagrangian describing the interactions of the SM--like Higgs scalar $h$
and $Z$--boson with $\chi_1$ and $\chi_2$ is given by
\begin{equation}
\mathcal{L}_{Zh\chi}=
\sum_{\alpha,\sigma} \dfrac{M_Z}{2 v} R_{Z\alpha\sigma} Z_{\mu} \biggl(\chi^{T}_{\alpha}\gamma_{\mu}\gamma_{5}\chi_{\sigma}\biggr) +
\sum_{\alpha,\sigma} g_{h\alpha\sigma} h \biggl(\chi^{T}_{\alpha} \chi_{\sigma}\biggr)\,,
\label{19}
\end{equation}
$$
R_{Z\alpha\sigma}=N_{\alpha}^1 N_{\sigma}^1 - N_{\alpha}^2 N_{\sigma}^2\,,\qquad\qquad
g_{h\alpha\sigma}= -\dfrac{1}{\sqrt{2}}\Biggl( f_{11} N^3_{\alpha} N^2_{\sigma} \cos\beta +
\tilde{f}_{11} N^3_{\alpha} N^1_{\sigma} \sin\beta \Biggr)\,,
$$
where $\alpha,\sigma=1,2$. The entries of the mixing matrix $N^a_i$ can be computed using the perturbation theory method.
Substituting the corresponding analytical expressions into Eqs.~(\ref{19}) one obtains the approximate formulae
\begin{equation}
|g_{h11}|\simeq \dfrac{\Delta_1}{v}\,,\qquad\qquad
R_{Z11}\simeq \dfrac{v^2(f_{11}^2\cos^2\beta - \tilde{f}_{11}^2 \sin^2\beta)}{4 \mu_{11} (\widetilde{\mu}_1 - \mu_{11})}\,.
\label{20}
\end{equation}
which are valid when $\widetilde{\mu}_1\gg \mu_{11}>0$ and $\mu_{11}$ is considerably larger than
$\tilde{f}_{11} v \sin\beta$ and $f_{11} v \cos\beta$.

The couplings of the SM--like Higgs scalar $h$ with mass $m_h\simeq 125\,\mbox{GeV}$ to nucleons $g_{hNN}$ is set by
\begin{equation}
g_{hNN} = a^{N}_{S} \dfrac{m_N}{v}\,,\qquad\qquad\qquad
a^{N}_{S} = \sum_{q=u,d,s} f^N_{Tq} + \dfrac{2}{27} \sum_{Q=c,b,t} f^N_{TQ}\,,
\label{21}
\end{equation}
where $N=p,n$, $m_N$ is a nucleon mass and
\begin{equation}
\langle N | m_{q}\bar{q}q |N \rangle= m_N f^N_{Tq}\,, \qquad\qquad\qquad\qquad f^N_{TQ} = 1 - \sum_{q=u,d,s} f^N_{Tq}\,.
\label{22}
\end{equation}
From Eqs.~(\ref{21}) and (\ref{22}) it follows that the couplings $g_{hNN}$ are determined by
the hadronic matrix elements, i.e., the coefficients $f^N_{Tq}$. These coefficients are related to the $\pi$--nucleon
$\sigma$ term and the spin content of the nucleon. In our analysis we set $f^p_{Tq}\simeq f^n_{Tq}\simeq f_{Tq}$,
so that $a^{p}_{S}\simeq a^{n}_{S}\simeq a_{S}$. Here we also fix $f_{Tu}\simeq 0.0153$, $f_{Td}\simeq 0.0191$ and
$f_{Ts}\simeq 0.0447$ which are the default values used in micrOMEGAs~\cite{Belanger:2013oya}.
The $t$-channel exchange of the lightest CP--even Higgs scalar gives rise to the spin--independent part
of $\chi_1$--nucleon cross section which is given by
\begin{equation}
\sigma_{SI} = \dfrac{4 m^2_r m_N^2}{\pi v^2 m^4_{h_1}} |g_{h \chi \chi} a_S|^2 \,,\qquad\qquad\qquad m_r = \dfrac{m_{\chi_1} m_N}{m_{\chi_1} + m_N}\,.
\label{23}
\end{equation}

The interactions of the $Z$--boson with quarks are described by
\begin{equation}
\mathcal{L}_{Zq} = \sum_q \dfrac{\bar{g}}{2}\, \overline{q} (a^{q}_{V} \gamma^{\mu} + a^{q}_{PV} \gamma^{\mu} \gamma^{5})q \, Z_{\mu}
=\dfrac{\bar{g}}{2} J^{\mu}_{NC} Z_{\mu}\,,
\label{24}
\end{equation}
$$
a^{q}_{V}=T_{3q} - 2 s^2_W Q_q\,,\qquad\qquad a^{q}_{PV}=T_{3q}\,.
$$
Using the Lagrangian (\ref{24}) the following hadronic matrix elements can be computed
\begin{equation}
\langle N'|J_{ NC}^{\mu}|N \rangle=\bar{\psi}'_N\gamma^{\mu}\left(a_V^N+a_{PV}^N\gamma_5\right)\psi_N\,,
\label{25}
\end{equation}
\begin{equation}
a_V^p\simeq\left(\frac{1}{2}-2s^2_W\right),\qquad
a_{PV}^p\simeq\left(\frac{1}{2}\Delta^{(p)}_u-\frac{1}{2}\Delta^{(p)}_d-\frac{1}{2}\Delta^{(p)}_s\right)\,,
\label{26}
\end{equation}
\begin{equation}
a_V^n\simeq -\frac{1}{2},\qquad
a_{PV}^n\simeq \left(\frac{1}{2}\Delta^{(n)}_u-\frac{1}{2}\Delta^{(n)}_d-\frac{1}{2}\Delta^{(n)}_s\right)\,,
\label{27}
\end{equation}
where $\Delta^{p}_u=\Delta^{n}_d=0.777$, $\Delta^{p}_d=\Delta^{n}_u=-0.438$ and $\Delta^{p}_s=\Delta^{n}_s=-0.053$ \cite{Lin:2018obj}.
From Eqs.~(\ref{26}) and (\ref{27}) one finds $a_{PV}^p \simeq 0.63$ and $a_{PV}^n \simeq -0.58$.
The $t$-channel exchange of the $Z$--boson results in the spin--dependent parts
of $\chi_1$--proton and $\chi_1$--neutron cross sections
\begin{equation}
\sigma^{p}=\dfrac{3 m^2_r}{\pi v^4} |R_{Z11} a^{p}_{PV}|^2\,,\qquad\qquad\qquad
\sigma^{n}=\dfrac{3 m^2_r}{\pi v^4} |R_{Z11} a^{n}_{PV}|^2\,.
\label{28}
\end{equation}

For each set of parameters $\tan\beta$, $\mu_{11}$, $\widetilde{\mu}_1$, $f_{11}$ and $\tilde{f}_{11}$
the $\chi_1$--nucleon cross sections $\sigma_{SI}$, $\sigma^{p}$ and $\sigma^{n}$ may be calculated.
The value of $\tan\beta$ in the SE$_6$SSM is less constrained as compared with the MSSM.
In particular, in the MSSM the scenarios with moderate values of $\tan\beta$, i.e. $\tan\beta\lesssim 4$ are
ruled out because the lightest Higgs scalar has a mass which is considerably smaller than $125\,\mbox{GeV}$.
In the SE$_6$SSM with $\lambda\gtrsim \sqrt{2} (M_Z/v)\simeq 0.5$ one can find the scenarios
with $125\,\mbox{GeV}$ SM--like Higgs for any $\tan\beta\gtrsim 2$. For so large values of $\lambda$
all Higgs bosons except the lightest CP--even Higgs scalar tend to have masses beyond the multi-TeV range
so that they cannot be discovered at the LHC \cite{King:2005jy,King:2005my,King:2006vu,King:2006rh}.
To simplify our analysis we set $\tan\beta\simeq 3$ and $\widetilde{\mu}_1\simeq 2\,\mbox{TeV}$.

The parameter space in the $E_6$ inspired $U(1)$ extensions of the MSSM is strongly constrained by the LHC
experimental lower bounds on the $Z'$ masses. Such gauge bosons are required to be heavier than
$4.5\,\mbox{TeV}$ \cite{CMS:2021ctt,ATLAS:2019erb}. For $\langle S \rangle \simeq \langle \overline{S} \rangle$
the $Z'$ mass in the SE$_6$SSM is given by
\begin{equation}
M_{Z'}\approx 2 g^{\prime}_1 Q_S \,\langle S \rangle\,.
\label{29}
\end{equation}
Assuming gauge coupling unification one can compute the low--energy value of $g^{\prime}_1$. With this value of $g^{\prime}_1$
the VEVs $\langle S \rangle \simeq \langle \overline{S} \rangle$ have to be larger than $6\,\mbox{TeV}$ to ensure that
$M_{Z'}\gtrsim 4.5\,\mbox{TeV}$.

\begin{figure}[!htb]
\centering
\hfill
\includegraphics[width=0.495\linewidth]{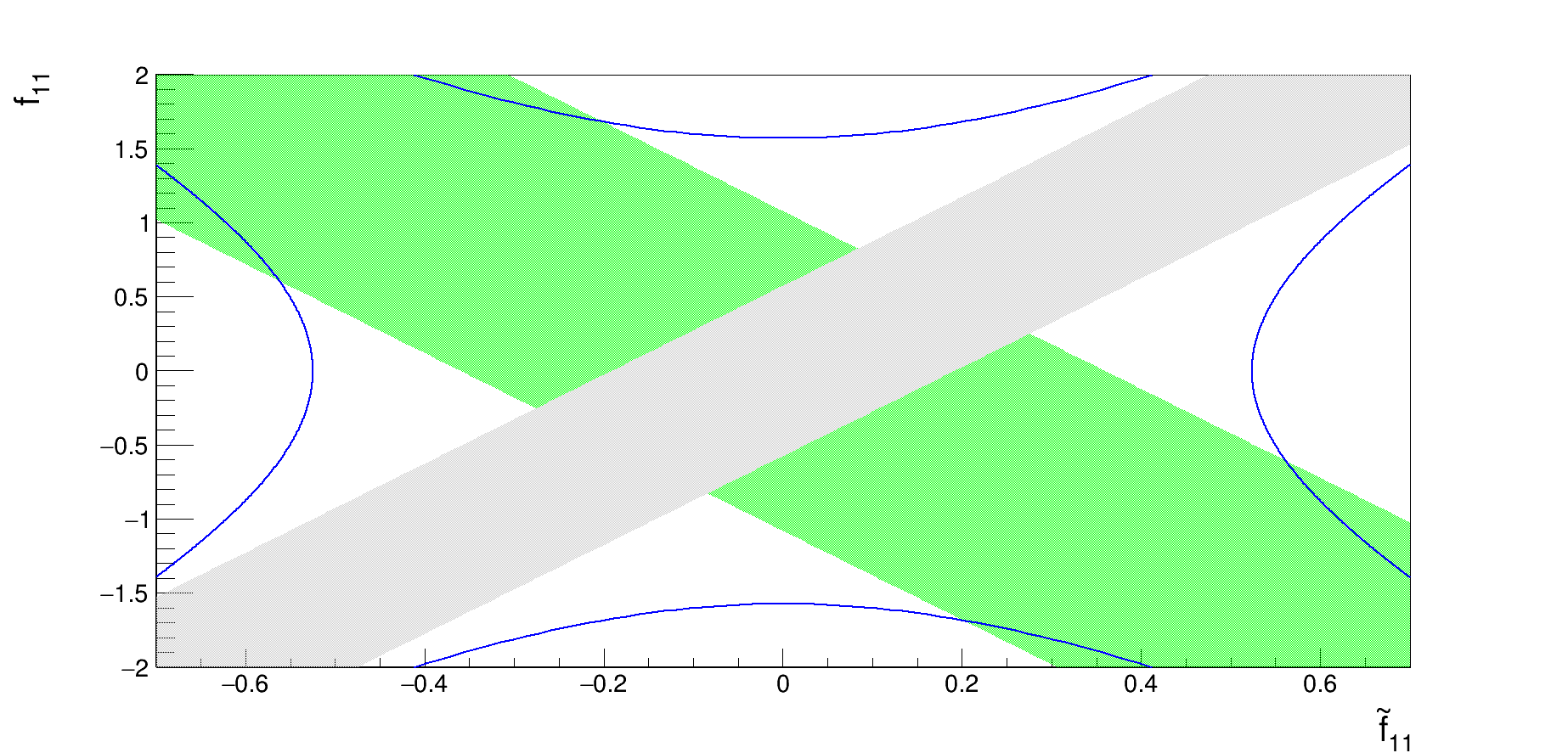}
\includegraphics[width=0.495\linewidth]{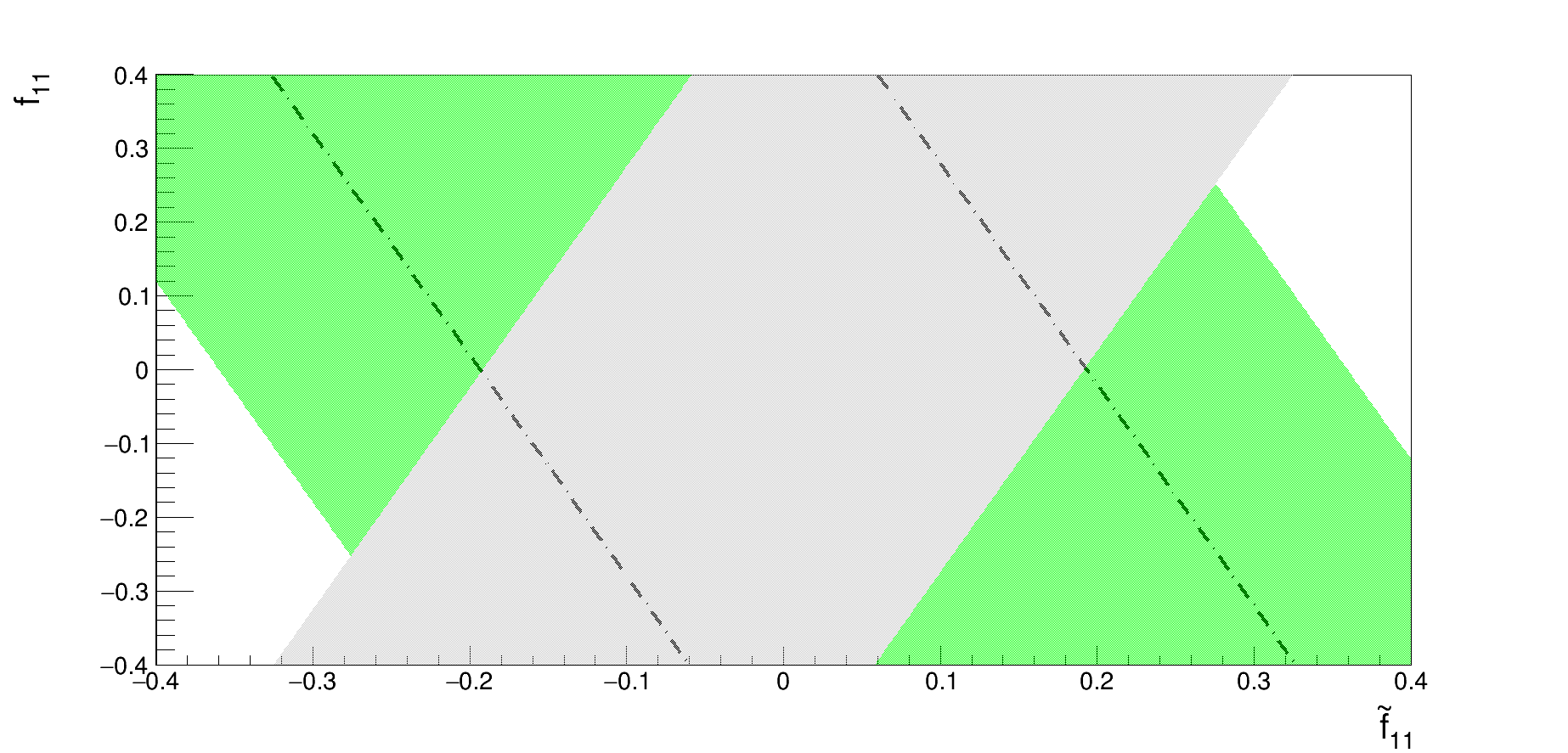}
\hfill
\caption{The constraints on the SE$_6$SSM parameter space in the $\tilde{f}_{11} - f_{11}$ plane
for $\tan\beta=3$, $\tilde{\mu}_1=2\,\mbox{TeV}$ and $\mu_{11}=500\,\mbox{GeV}$.
Green region marks the part of the parameter space, where the experimental restriction on
$\sigma_{SI}$ is satisfied. In the grey area $\Delta_2\le 200\,\mbox{MeV}$.
({\bf Left}) The area limited by the solid lines corresponds to the region
where $\sigma^{n} < \sigma^{n}_0/10$. ({\bf Right}) The dashed--dotted lines
limit the part of the parameter space, where $\sigma_{SI} < \sigma^0_{SI}/10$.}
\label{fig1}
\end{figure}

The evaluated $\chi_1$--nucleon cross sections $\sigma_{SI}$, $\sigma^{n}$ and
$\sigma^{p}$ have to be compared with the most stringent experimental bounds set
by the LZ \cite{LZ:2024zvo} and IceCube \cite{IceCube:2025fcu} recently. However one needs to take into
account that for $m_{\chi_1}\approx \mu_{11}\lesssim 1\,\mbox{TeV}$ the lightest exotic neutralino states compose only some
fraction of the cold dark matter density. Thereby, the experimental limits become weaker, i.e.
\begin{equation}
\sigma_{SI} < \sigma^0_{SI}=\frac{(\Omega h^2)_{\text{exp}}}{\Omega_{\tilde{H}} h^2} (\sigma_{SI})_{\text{exp}}\,,\qquad\qquad
\sigma^{p,n} < \sigma^{p,n}_0=\frac{(\Omega h^2)_{\text{exp}}}{\Omega_{\tilde{H}} h^2} (\sigma^{p,n})_{\text{exp}}\,.
\label{30}
\end{equation}
Here $\Omega_{\tilde{H}} h^2$, $\sigma_{SI}$ and $\sigma^{p,n}$ are calculated values of the cold dark matter density
and $\chi_1$--nucleon cross sections for each set of $\mu_{11}$, $f_{11}$ and $\tilde{f}_{11}$.
At the same time $(\sigma_{SI})_{\text{exp}}$ and $(\sigma^{p,n})_{\text{exp}}$
are the experimental bounds on the spin-independent $\chi_1$--nucleon cross section as well as
spin-dependent $\chi_1$--proton and $\chi_1$--neutron cross sections at the given mass $m_{\chi_1}$.
Any set of parameters, which does not satisfy the conditions (\ref{30}), is ruled out.

\begin{figure}[!htb]
\centering
\hfill
\includegraphics[width=0.495\linewidth]{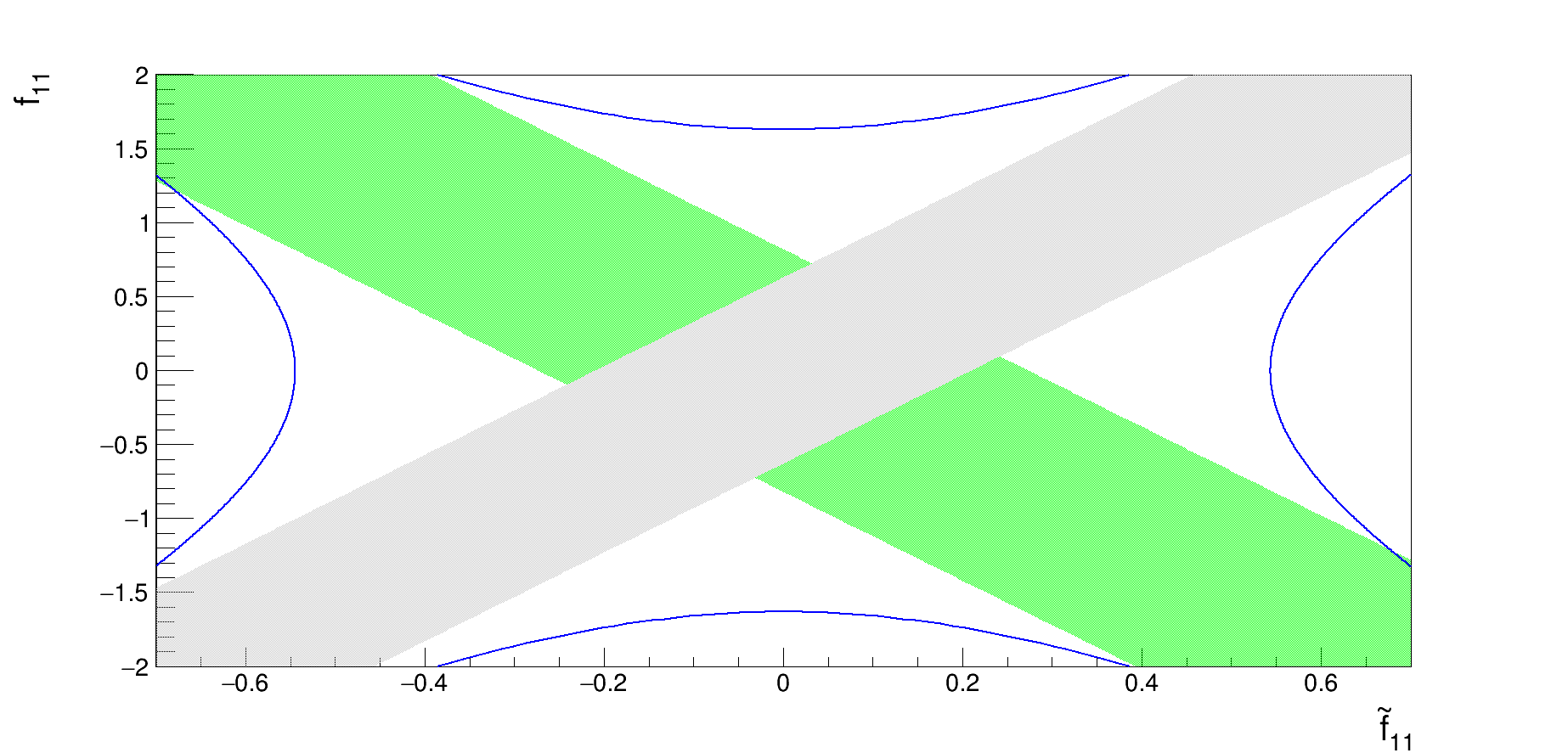}
\includegraphics[width=0.495\linewidth]{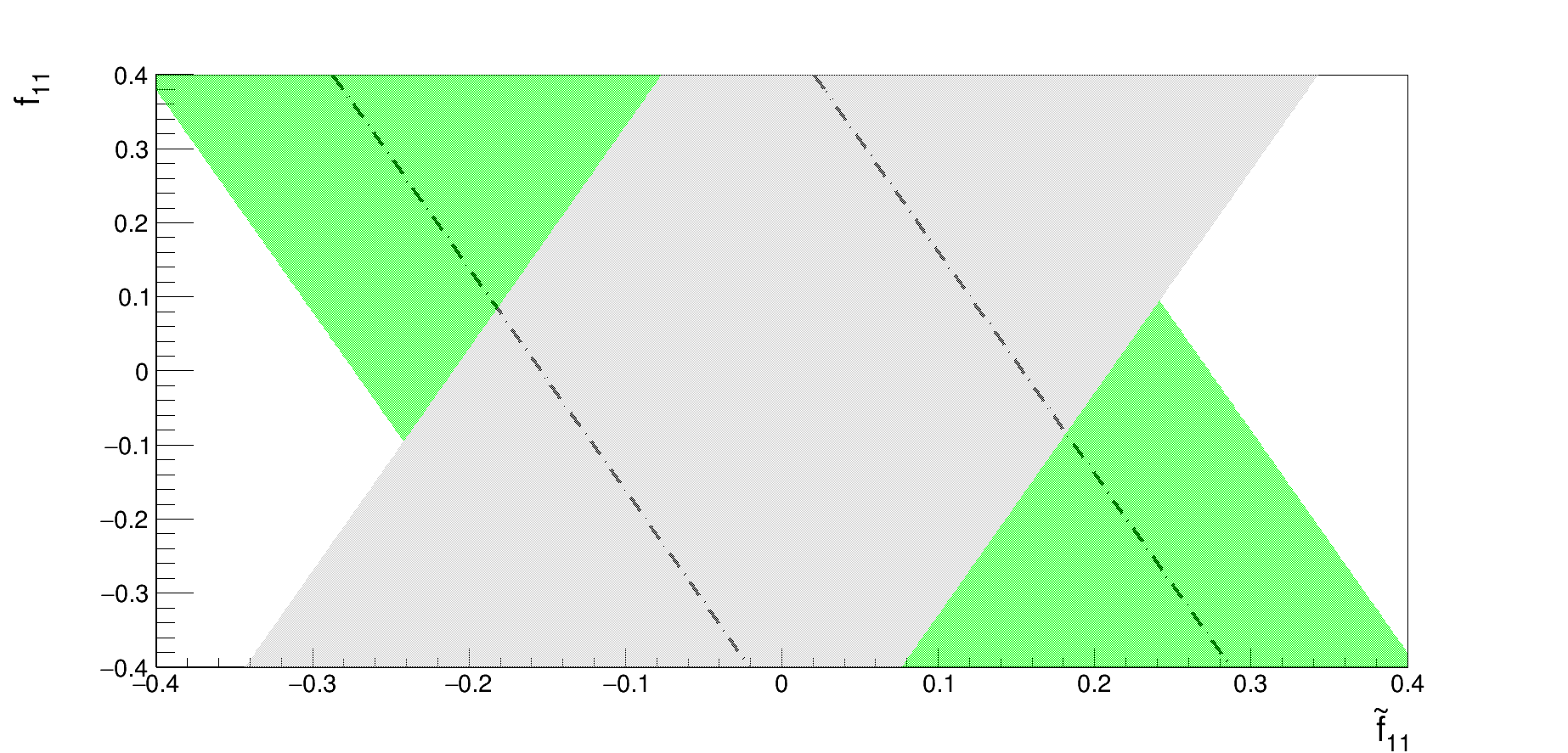}
\hfill
\caption{The constraints on the SE$_6$SSM parameter space in the $\tilde{f}_{11} - f_{11}$ plane
for $\tan\beta=3$, $\tilde{\mu}_1=2\,\mbox{TeV}$ and $\mu_{11}=1000\,\mbox{GeV}$.
Green region marks the part of the parameter space, where the experimental restriction on
$\sigma_{SI}$ is satisfied. In the grey area $\Delta_2\le 200\,\mbox{MeV}$.
({\bf Left}) The area limited by the solid lines corresponds to the region
where $\sigma^{n} < \sigma^{n}_0/10$. ({\bf Right}) The dashed--dotted lines
limit the part of the parameter space, where $\sigma_{SI} < \sigma^0_{SI}/10$.}
\label{fig2}
\end{figure}

The results of our analysis are presented in Figs.~\ref{fig1} and \ref{fig2}.
We consider two scenarios associated with $\mu_{11}\simeq 500\,\mbox{GeV}$ and $\mu_{11}\simeq 1\,\mbox{TeV}$.
Eqs.~(\ref{15}), (\ref{20}), (\ref{23}) and (\ref{28}) indicate that for fixed values of
$\tan\beta$, $\widetilde{\mu}_1$ and $\mu_{11}$ the conditions (\ref{30}) set limits on the Yukawa couplings
$f_{11}$ and $\tilde{f}_{11}$. Therefore in Figs.~\ref{fig1} and \ref{fig2} the allowed regions in the
$\tilde{f}_{11} - f_{11}$ plane are shown. To guarantee the validity of perturbation theory up to the
GUT scale $M_X$ we require the absolute values of $f_{11}$ and $\tilde{f}_{11}$ to be smaller than $0.4$.

As follows from Eqs.~(\ref{28}) the computed values of $\sigma^{p}$ and $\sigma^{n}$ are always relatively close.
On the other hand the experimental limits $\sigma^{p}_{\text{exp}}$ and $\sigma^{n}_{\text{exp}}$ are very different.
Taking into account that the lightest exotic neutralino states annihilate mainly into a pair of gauge bosons, 
the experimental bounds on the spin--dependent WIMP--proton scattering cross section
$\sigma^{p}_{\text{exp}}\approx 6\cdot 10^3\,\mbox{zb}\, (10^4\,\mbox{zb})$ for $m_{\chi_1}\approx 500\,\mbox{GeV} (1\,\mbox{TeV})$,
where $1\,\mbox{zb}=10^{-45}\,\mbox{cm}^2$ \cite{IceCube:2025fcu}. The spin--dependent WIMP--neutron scattering cross
section is more tightly constrained, i.e. $\sigma^{n}_{\text{exp}}\approx 2300 \,\mbox{zb}\, (5000 \,\mbox{zb})$
for $m_{\chi_1}\approx 500\,\mbox{GeV} (1\,\mbox{TeV})$ \cite{LZ:2024zvo}. These experimental bounds set rather weak constraints on
the Yukawa couplings $f_{11}$ and $\tilde{f}_{11}$. In Figs.~\ref{fig1} and \ref{fig2} the regions of the parameter space,
where $\sigma^{n} < \sigma^{n}_0/10$, are limited by the solid lines.

The most stringent restrictions on the Yukawa couplings $f_{11}$ and $\tilde{f}_{11}$ come from experimental bounds
on the spin--independent WIMP--nucleon scattering cross section. The corresponding experimental limits are $12\,\mbox{yb}$
for $m_{\chi_1}\approx 500\,\mbox{GeV}$ and $30\,\mbox{yb}$ for $m_{\chi_1}\approx 1\,\mbox{TeV}$ ($1\,\mbox{yb}=10^{-48}\,\mbox{cm}^2$)
which are the values of $(\sigma_{SI})_{\text{exp}}$ obtained by the LZ experiment \cite{LZ:2024zvo}.
The green areas in Figs.~\ref{fig1} and \ref{fig2} represent the allowed ranges of the Yukawa couplings $f_{11}$ and $\tilde{f}_{11}$
where $\sigma_{SI} < \sigma^0_{SI}$. The requirement $\Delta_2\le 200\,\mbox{MeV}$ disfavours the grey regions.
The areas between dashed--dotted lines correspond to the regions of the parameter space, where $\sigma_{SI} < \sigma^0_{SI}/10$.
Thus even if the experimental bounds become considerably more stringent the scenarios under consideration are not going to be ruled out.


\section{Conclusions}

The cancellation of anomalies in the $E_6$ inspired $U(1)$-extended SUSY models requires that the low energy matter
content of these models involves three fundamental $27$ representations of $E_6$. Three $27$-plets, in particular,
includes three SM singlet superfields $S_i$ and three families of Higgs-like doublets $H^{u}_{i}$ and $H^{d}_{i}$.
One pair of such Higgs--like supermultiplets ($H_d\equiv H^{d}_{3}$ and $H_u\equiv H^{u}_3$) acquires VEVs
breaking the EW symmetry. The fermion components of two other families $H^{u}_{\alpha}$ and $H^{d}_{\alpha}$
($\alpha=1,2$) as well as the fermion components of the SM singlet superfields $S_i$, which do not gain VEVs,
compose exotic neutralino and chargino states. The lightest exotic neutralino in these $U(1)$ extensions of the MSSM
can be stable forming some part of the cold dark matter density.

In this article we examine the constraints on the couplings of such lightest exotic neutralino
within a specific extension of the MSSM with extra $U(1)_N$ gauge symmetry (SE$_6$SSM), in which
the single discrete $\tilde{Z}^{H}_2$ symmetry forbids the most dangerous baryon and lepton number
violating operators as well as tree-level non-diagonal flavor transitions.
In addition to three $27$-plets the low energy spectrum of the SE$_6$SSM is supplemented by four
$E_6$ singlet superfields, a pair of $SU(2)_W$ lepton doublets $L_4$ and $\overline{L}_4$ with
opposite $SU(2)_W \times U(1)_Y \times U(1)_N$ quantum numbers and a pair of the SM singlet superfields
$S$ and $\overline{S}$ with opposite $U(1)_N$ charges. The superfields $S$ and $\overline{S}$ can
acquire very large VEVs ($\langle S\rangle \simeq \langle \bar{S}\rangle \gg 10\,\mbox{TeV}$)
resulting in the breakdown of the $U(1)_N$ gauge symmetry and inducing sufficiently large masses of
all extra exotic particles. The supermultiplets $L_4$ and $\overline{L}_4$ facilitate the
gauge coupling unification and permit the lightest exotic quark/squark to decay before BBN.

The conservation of the $\tilde{Z}^{H}_2$ symmetry and $R$--parity in the SE$_6$SSM leads to
two stable neutral states which can form cold dark matter density. Here we assume that
one of these stable states is gravitino with mass $m_{3/2}\sim 1\,\mbox{GeV}$ while
another stable particle is the lightest exotic neutralino $\chi_1$ composed of the fermion
components of $H^{d}_{1}$ and $H^{u}_{1}$. In this case the lightest exotic chargino $\chi^{\pm}_1$
as well as the lightest and second lightest exotic neutralinos ($\chi_1$ and $\chi_2$) are nearly
degenerate. If the masses of these exotic particles are smaller than $1.1\,\mbox{TeV}$ they may
result in the phenomenologically acceptable dark matter density.

The interactions of dark matter with the nucleons in the SE$_6$SSM are determined by the couplings of $\chi_1$.
In this article we focused on the scenarios in which all exotic fermions and all scalars except the SM--like Higgs,
$\chi_1$, $\chi_2$, $\chi^{\pm}_1$ and gravitino are heavier than $10\,\mbox{TeV}$. As a consequence
the main contributions to the spin--dependent $\chi_1$--proton and $\chi_1$--neutron cross-sections ($\sigma^{p}$ and $\sigma^{n}$)
come from the diagrams with t--channel exchange of the $Z$--boson whereas the spin--independent
$\chi_1$--nucleon cross-section $\sigma_{SI}$ is dominated by the $t$-channel exchange of the SM--like Higgs scalar.
Our analysis revealed that there is a significant part of the SE$_6$SSM parameter space where $\sigma_{SI}$ and
$\sigma^{p,n}$ are considerably lower than the present experimental bounds.

The phenomenological viability of the SE$_6$SSM requires $\chi^{\pm}_1$, $\chi_2$ and $\chi_1$ to be lighter than $1.1\,\mbox{TeV}$. 
Otherwise the contribution of the lightest exotic neutralino to the total dark matter density becomes larger than its measured value. 
If $\chi_1$ is considerably lighter than $1.1\,\mbox{TeV}$ its annihilation cross section is sufficiently large giving rise to 
the relatively small density of $\chi_1$. In this case one can naively expect that the indirect signal from dark matter annihilation 
may be enhanced. However the analysis of the similar scenarios performed within the MSSM indicates that the corresponding indirect 
signal gets weaker with diminishing of the lightest neutralino mass because the density of $\chi_1$ decreases \cite{Baer:2018rhs}.

Since scenarios under consideration imply that $\chi^{\pm}_1$, $\chi_2$ and $\chi_1$ are lighter than $1.1\,\mbox{TeV}$
they could be produced at the LHC. Nevertheless when the masses of the lightest exotic chargino $m_{\chi^{\pm}_1}$,
lightest and second lightest exotic neutralino ($m_{\chi_1}$ and $m_{\chi_2}$) are rather close, the decay products of
$\chi_2$ and $\chi^{\pm}_1$ may escape detection. This also occurs within natural SUSY if the mass splitting between the
lightest chargino and neutralino states is a few GeV or even smaller \cite{Baer:2012up,Baer:2012uy,Baer:2012cf}.
The results of the searches in this case depend on the values $\Delta_0 = m_{\chi_2}-m_{\chi_1}$ and $\Delta = m_{\chi^{\pm}_1}-m_{\chi_1}$.
When $\Delta \simeq 2\,\mbox{GeV} (4.7\,\mbox{GeV})$ ATLAS ruled out the lightest chargino with mass below $140\,\mbox{GeV}
(193\,\mbox{GeV})$ \cite{ATLAS:2019lng} whereas CMS excluded the lightest chargino masses below $112\,\mbox{GeV}$
for $\Delta = 1\,\mbox{GeV}$ \cite{CMS:2019san}. For $\Delta \lesssim 150\,\mbox{MeV}$ the lightest chargino may be long--lived
and LHC experiments ruled out such charginos if they are lighter than $1090\,\mbox{GeV}$ \cite{ATLAS:2019gqq}.

The last most stringent experimental limit is not applicable because in the SE$_6$SSM scenarios under consideration
$\Delta\gtrsim 300\,\mbox{MeV}$ \cite{Nagata:2014wma,Cirelli:2005uq}. The second lightest exotic neutralino $\chi_2$ cannot be
long--lived as well. Indeed, at the tree level $\Delta_0 = \Delta_1 + \Delta_2 \gtrsim \Delta_2$ whereas in our analysis
we require $\Delta_2\ge 200\,\mbox{MeV}$. Thus the lightest exotic chargino and the second lightest exotic neutralino
decay into $\chi_1$ and hadrons. Since $\chi^{\pm}_1$, $\chi_2$ and $\chi_1$ are nearly degenerate
it seems to be rather problematic to discover this set of states at hadron colliders. The discovery prospects for such
exotic fermions look more promising at future International Linear Collider.

\section*{Conflicts of Interest}

The authors declare no conflicts of interest.


\end{document}